# Plump Cutthroat Trout and Thin Rainbow Trout in a Lentic Ecosystem

Joshua Courtney,[1] Jessica Abbott,[2] Kerri Schmidt,[2] and Michael Courtney[2]

[1]BTG Research, P.O. Box 62541, Colorado Springs, CO, 80962

[2]United States Air Force Academy, 2354 Fairchild Drive, USAF Academy, CO, 80840

Michael_Courtney@alum.mit.edu

**Background:** Much has been written about introduced rainbow trout (*Oncorhynchus mykiss*) interbreeding and outcompeting cutthroat trout (*Oncorhynchus clarkii*). However, the specific mechanisms by which rainbow trout and their hybrids outcompete cutthroat trout have not been thoroughly explored, and the published data is limited to lotic ecosystems.
**Materials and Methods:** Samples of rainbow trout and cutthroat trout were obtained from a lentic ecosystem by angling. The total length and weight of each fish was measured and the relative weight of each fish was computed (Anderson R.O., Neumann R.M. 1996. Length, Weight, and Associated Structural Indices, Pp. 447-481. In: Murphy B.E. and Willis D.W. (eds.) Fisheries Techniques, second edition. American Fisheries Society.), along with the mean and uncertainty in the mean for each species. Data from an independent source (K.D. Carlander, 1969. Handbook of Freshwater Fishery Biology, Volume One, Iowa University Press, Ames.) was also used to generate mean weight-length curves, as well as $25^{th}$ and $75^{th}$ percentile curves for each species to allow further comparison.
**Results:** The mean relative weight of the rainbow trout was 72.5 (+/- 2.1); whereas, the mean relative weight of the cutthroat trout was 101.0 (+/- 4.9). The rainbow trout were thin; 80% weighed below the $25^{th}$ percentile. The cutthroat trout were plump; 86% weighed above the $75^{th}$ percentile, and 29% were above the heaviest recorded specimens at a given length in the Carlander (1969) data set.
**Conclusion:** This data casts doubt on the hypothesis that rainbow trout are strong food competitors with cutthroat trout in lentic ecosystems. On the contrary, in the lake under study, the cutthroat trout seem to be outcompeting rainbow trout for the available food.

## Introduction
Since the late 1800s, most subspecies of cutthroat trout (*Oncorhynchus clarkii*) have experienced dramatic reductions in abundance and distribution, with the greenback cutthroat (*O. clarkii stomias*) being listed as threatened under the U.S. Endangered Species Act. (McGrath and Lewis 2007, Henderson et al. 2000) Many authors cite introduced rainbow trout (*Oncorhynchus mykiss*) as having a great impact on native cutthroat trout through hybridization and competition. (Henderson et al. 2000, Seiler and Keeley 2009, Quist and Hubert 2004, Behnke 2002) Some studies have focused on interspecies competition in lotic ecosystems (Seiler 2007, Seiler and Keeley 2007a, Seiler and Keeley 2007b, Fuller et al. 2012, Blinn et al. 1993), but little or no work has been published on competition in lentic ecosystems. This paper presents data showing that stocked cutthroat trout are competing well with stocked rainbow trout in a lentic ecosystem.

## Method
Small samples of rainbow trout (15) and cutthroat trout (7) were obtained by angling from Dead Man's Lake, a lentic ecosystem in El Paso county, Colorado in Spring 2010. The small lake was created as an impoundment of Dead Man's creek at an elevation near 7300 feet and is between 1 and 2 surface acres in size with a maximum depth of 9 feet. The total length and weight of each fish were measured and the relative weight of each fish was computed (Anderson and Neumann 1996), along with the mean and uncertainty in the mean relative weight for each species. The relative weight is 100 times the actual weight divided by the standard weight, the standard weight being the expected 75$^{th}$ percentile weight at a given length.

To allow further comparison with historical weight-length expectations, data from Carlander (1969) were used to generate mean weight-length curves, as well as 25$^{th}$ and 75$^{th}$ percentile curves for each species. Minimum and maximum curves were also generated from the columns in the Carlander data listing the minimum and maximum fish for each 25mm length interval. Non-linear regression applied to an improved weight-length model, $W(L) = (L/L_1)^b$ was used, where W is the weight in kg, L is the length in cm, b is the adjustable parameter exponent, and $L_1$ is a parameter corresponding to the typical length of a fish weighing 1 kg. This improved model has been shown to have much lower parameter uncertainties and lower covariance between parameters than the traditional weight-length model (Dexter et al. 2011, Keenan et al. 2011, Cole-Fletcher et al. 2011) and also overcomes the weaknesses associated with negative correlation between parameters in the approaches using linear regression. (Gerow et al. 2005) Given the number of fish in each interval, the minimum curve is likely between the 0$^{th}$ and 5$^{th}$ percentile, and the maximum curve is likely between the 95$^{th}$ and 100$^{th}$ percentile. To reduce the possibility of length related bias, only the Carlander data from 150mm to 400mm total length was use to produce the comparison curves.

The cutthroat trout did not have a range for the central 50% of weights listed in Carlander. It was observed that in rainbow trout, brook trout, and brown trout the width of the central 50% averaged 22.3%, 26.4%, and 26.6% of the extreme spread, with uncertainties of 3.1%, 3.6%, and 2.8%. Consequently, in cutthroat trout, the range of the central 50% was estimated as 25% of the extreme spread for each length interval. Once the minimum, 25$^{th}$ percentile, mean, 75$^{th}$ percentile, and maximum weight-length curves were generated for each species, the weight vs. total length data from the present study was plotted along with these curves.

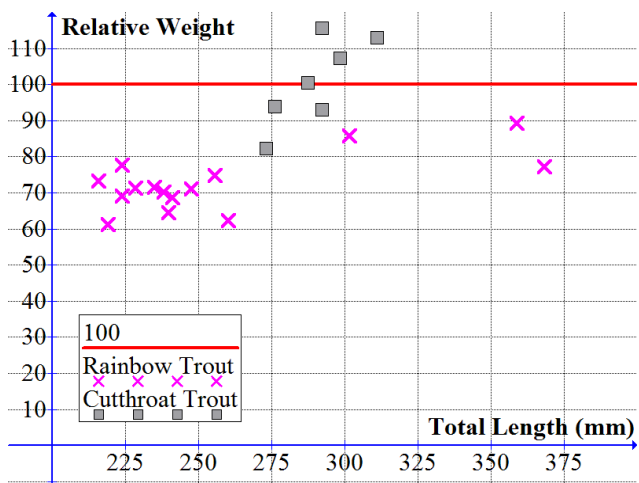

*Figure 1: Relative weights (Anderson and Neumann 1996) of cutthroat trout and rainbow trout from Dead Man's Lake (El Paso County, Colorado).*

**Results**
Figure 1 shows that the relative weights of the rainbow trout range from 61 to 89 with a mean of 72.5 (+/- 2.1), suggesting that the rainbow trout are thin. In contrast, the relative weights of the cutthroat trout range from 82 to 113, with a mean relative weight of 101.0 (+/- 4.1), showing that the cutthroat trout are plump.

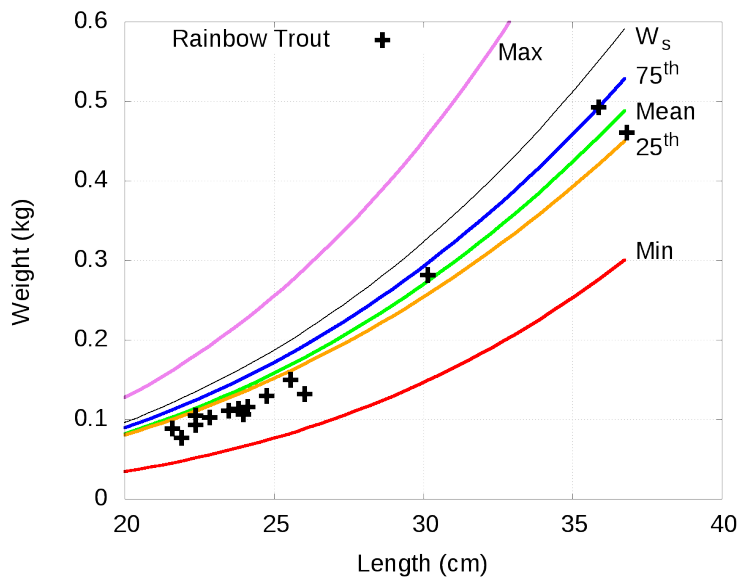

*Figure 2: Weight vs. length data for rainbow trout from Dead Man's Lake (El Paso County, Colorado) compared with the minimum, 25th percentile, mean, 75th percentile, and maximum best fit curves determined from Carlander (1969). The standard weight curve, $W_s$, (Anderson and Neumann 1996) is also shown for comparison.*

Figure 2 shows that 12 of 15 rainbow trout sampled were below the 25th percentile curve as determined from the Carlander (1969) data. These include all the fish under 30 cm in length; whereas, the three specimens above 30 cm in length tend to fall between the 25th and 75th percentile curves. It is also notable that the standard weight curve (Anderson and Neumann 1996) is larger than the 75th percentile curve from Carlander over the range of interest. Table 1 shows the weight-length parameters corresponding to the curves in Figure 2. It is expected that $L_1$, the length of a fish weighting 1 kg, decreases sequentially from the minimum to the maximum curves.

*Table 1: Best fit parameters, uncertainties, and correlation coefficients, r, for curves generated from Carlander (1969) data for rainbow trout.*

|      | b     | uncertainty % | $L_1$ cm | uncertainty % | r     |
|------|-------|---------------|----------|---------------|-------|
| Min  | 3.548 | 7.904         | 51.531   | 3.019         | 0.989 |
| 25th | 2.827 | 2.189         | 48.706   | 0.786         | 0.999 |
| Mean | 2.927 | 2.109         | 46.918   | 0.675         | 0.999 |
| 75th | 2.915 | 3.064         | 45.720   | 0.911         | 0.998 |
| Max  | 3.112 | 6.334         | 38.754   | 1.011         | 0.992 |

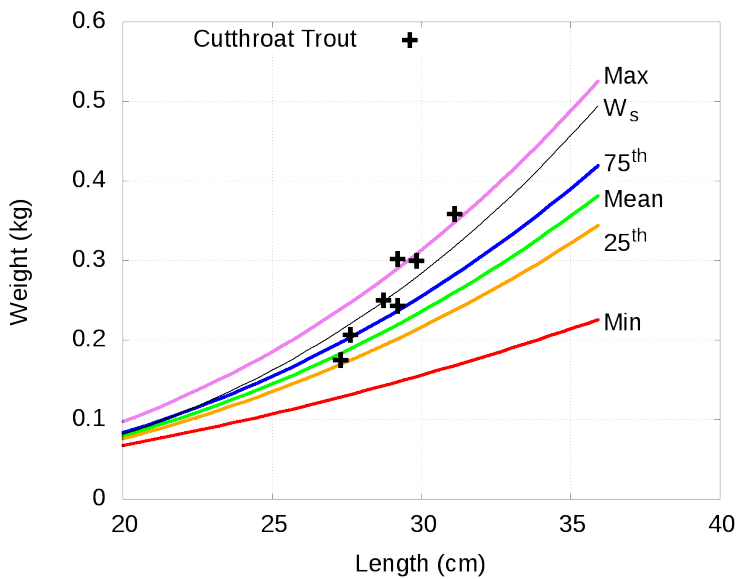

*Figure 3: Weight vs. length data for cutthroat trout from Dead Man's Lake (El Paso County, Colorado) compared with the minimum, 25th percentile, mean, 75th percentile, and maximum best fit curves determined from Carlander (1969). The standard weight curve, $W_s$, (Anderson and Neumann 1996) is also shown for comparison.*

Figure 3 shows that six of the seven cutthroat trout sampled were above the 75th percentile curve as determined from the Carlander (1969) data. Two of seven were above the maximum weight compiled by Carlander (1969) for their length. Clearly, these specimens are competing well for food. It is also notable that the standard weight curve (Anderson and Neumann 1996) is larger than the 75th percentile curve from Carlander over the range of interest. Table 2 shows the weight-length parameters corresponding to the curves in Figure 3. It is expected that $L_1$, the length of a fish weighting 1 kg, decreases sequentially from the minimum to the maximum curves. It was not expected that all the values of the exponent, b, would be less than 3, or that they would increase sequentially from the minimum to the maximum curve.

Plump Cutthroat Trout and Thin Rainbow Trout in a Lentic Ecosystem*Table 2: Best fit parameters, uncertainties, and correlation coefficients, r, for curves generated from Carlander (1969) data for cutthroat trout.*

|  | b | uncertainty % | $L_1$ cm | uncertainty % | r |
|---|---|---|---|---|---|
| Min | 2.056 | 14.170 | 74.077 | 11.500 | 0.957 |
| 25th | 2.576 | 2.792 | 54.368 | 1.187 | 0.997 |
| Mean | 2.672 | 2.377 | 51.521 | 1.006 | 0.999 |
| 75th | 2.756 | 1.770 | 49.245 | 0.666 | 0.999 |
| Max | 2.876 | 4.343 | 44.917 | 1.248 | 0.996 |

**Discussion**

The results show that the cutthroat trout in Dead Man's Lake seem to be acquiring more than their share of the available food, leaving the rainbow trout in poor condition, especially the rainbow trout below 30 cm long.  Since the fish were stocked in late winter, it may be that the cutthroat trout simply adapt to the foraging conditions of the lake more quickly that the rainbow trout.  In contrast, stocked rainbow trout in similarly sized lentic ecosystems in El Paso County without competing cutthroat trout have much higher relative weights.  For example, 28 rainbow trout sampled from Lower Twin Pond (Monument, CO) had a mean relative weight of 107.5 (+/- 2.3), and 10 rainbow trout sampled from Hidden Pond (Monument, CO) had a mean relative weight of 100.4 (+/- 1.6).

Displacement of native cutthroat trout by rainbow trout and introgression of rainbow trout genetics into cuttroat trout populations are widely recognized as significant problems.  (Wiegel et al. 2003)  Cutthroat trout have been recognized to be strong competitors at higher elevations, often leading to a gradual shift from mixed to more pure genetics as one moves progressively toward the headwaters of a given lotic system. (Wiegel et al. 2003)  One prior study also found that cutthroat trout grow faster and larger than rainbow trout when the two are in sympatry in British Columbia lakes. (Nilsson and Northcote 1981)  If the results of the present study are found to be more broadly applicable, and cutthroat trout are generally strong food competitors with rainbow trout in lentic ecosystems at lower elevations, then lentic ecosystems might prove useful for exerting selection pressure in favor of more pure cutthroat trout genetics at lower elevations and in systems where strongly mixed genetics are already present.  Due to the small samples and consideration of a single lentic system with competing species, the results of this study should not be considered definitive, but rather suggest the need to study the competition between rainbow and cutthroat trout in additional lentic ecosystems.

**Acknowledgements**
The authors acknowledge helpful discussions with Amy Courtney, PhD (BTG Research).  We are grateful to Lt Col Andy Gaydon (USAFA/DFMS) for reading the manuscript and offering helpful comments.  We also thank BTG Research and the Quantitative Reasoning Center at the United States Air Force Academy for supporting this work.5

**Plump Cutthroat Trout and Thin Rainbow Trout in a Lentic Ecosystem**Seiler S.M. 2007. Ecological and environmental investigations of competition between native Yellowstone cutthroat trout (*Oncorhynchus clarkii bouvieri*), rainbow trout (*Oncorhynchus mykiss*) and their hybrids. Ph.D. thesis, Idaho State University, Pocatello, Idaho, USA.

Seiler S.M., Keeley E.R. 2007a. Morphological and swimming stamina differences between Yellowstone cutthroat trout (*Oncorhynchus clarkii bouvieri*), rainbow trout (*Oncorhynchus mykiss*), and their hybrids. Can. J. Fish. Aquat. Sci. 64: 127–135.

Seiler S.M., Keeley E.R. 2007b. A comparison of aggressive and foraging behaviour between juvenile cutthroat trout, rainbow trout, and F1 hybrids. Anim. Behav. 74: 1805–1812.

Seiler S.M., Keeley E.R. 2009. Competition between native and introduced salmonid fishes: cutthroat trout have lower growth rate in the presence of cutthroat–rainbow trout hybrids. Can. J. Fish. Aquat. Sci. 66: 133–141.

Quist M.C., Hubert W.A. 2004. Bioinvasive species and the preservation of cutthroat trout in the western United States: ecological, social, and economic issues. Environmental Science & Policy 7:303-313.

Weigel D.E., PetersonJ.T., Spruell P. 2003. Introgressive Hybridization between Native Cutthroat Trout and Introduced Rainbow Trout. Ecological Applications 13:38–50.
7